\documentclass[12pt]{article}
\usepackage[pdftex]{graphicx}
\usepackage{amsmath}
\usepackage{xcolor,soul}
\usepackage{url}
\usepackage{amssymb,esint}

\title{\bf Non-cosmological, Non-Doppler Relativistic Frequency
  Shift over Astronomical Distances}
\author{Germano D'Abramo\\
{\small Ministero dell'Istruzione, dell'Universit\`a e della Ricerca,}\\
{\small 00041, Albano Laziale, RM, Italy}\\
{\small E--mail: {\tt germano.dabramo@gmail.com}}\\
{\small ORCID: 0000-0003-1277-7418}}

\date{{\em Dynamics} {\bf 2022}, 2(3), 319--325\\
  \vspace{0.5cm}
  \url{https://doi.org/10.3390/dynamics2030017}}

\begin{document}

\maketitle

\begin{abstract}
 We investigate in detail an apparently unnoticed consequence of
 special relativity. It consists in time dilation/contraction and frequency
 shift for emitted light affecting accelerated reference frames at astronomical
 distances from an inertial observer. The frequency shift is non-cosmological
 and non-Doppler in nature. We derive the main formulae and compare their
 predictions with the astronomical data available for Proxima Centauri.
 We found no correspondence with observations. Since the implications of the
 new time dilation/contraction and frequency shift are blatantly paradoxical,
 we do not expect to find one. By all indications, we are dealing with a
 genuine, and not a merely apparent, relativity paradox.\\
  
    
\end{abstract}

\section{Introduction}
\label{se1}

In the present paper, we focus on a consequence of special relativity that
seems to have gone unnoticed. It consists in time dilation/contraction
and consequent frequency shift for emitted light affecting accelerated
reference frames at astronomical distances from an inertial observer. For these
effects to show up, high relative velocities ($v\lesssim c$) are not necessary.
The frequency shift derived hereafter is purely relativistic. It has
nothing to do with the standard Doppler shift, which depends upon relative
speed, nor with the cosmological redshift.
These effects derive from a straightforward application of the Lorentz
transformation of the time coordinate
\begin{equation}
t' = \frac{t-\frac{vx}{c^2}}{\sqrt{1-\frac{v^2}{c^2}}},
\label{eq0}
\end{equation}
where $t'$ is the time coordinate of a frame moving with constant velocity
$v$ along the $x$-direction of a stationary inertial reference frame with
time coordinate $t$ (and parallel coordinate axes).
As usual, $c$ stands for the velocity of light.

The phenomenon described here has already been discussed in \cite{dab} and
is strictly related to the well-known Andromeda
paradox \cite{dab,rie,put,pen,wi}.

In the following two sections, we derive the time dilation/contraction and the
frequency shift formulae. {{Incidentally, they correspond to those
derived from general relativity in the case of a weak and spatially
uniform gravitational field.}}
We shall show that these formulae also hold when the
distant light source is inertial and stationary, and the observer accelerates. 

In Section \ref{se4}, we use these formulae to calculate the
expected frequency shift of the light emitted by Proxima Centauri and compare
it to the standard Doppler shift coming from the radial velocity imparted to
the star by orbiting planet Proxima Centauri b. The aim is to see whether there
are measurable consequences already for a relatively close astronomical source.

In Section \ref{se5}, we shall discuss the import of these new time
dilation/con\-trac\-tion and frequency shift phenomena. We show that despite
the straightforward derivation, the predicted effects on
the astronomical scale are loudly missing. Owing to the intrinsic paradoxical
implications of the derived time dilation/con\-trac\-tion and frequency shift
formulae, we expect not to find any observable effect.
The fascinating aspect is that such paradoxes appear to have a
genuine, and not a merely apparent, nature.

\section{Quick Derivation of Purely Relativistic Time Dilation/Contraction and Frequency Shift}
\label{se2}

Consider two reference frames, $S$ and $S'$. Initially, they are both
inertial and relatively at rest, with parallel coordinate axes. Frame $S$ is
the observer frame, while frame $S'$ is the frame of the light source.
Frame $S'$ is placed at an astronomical distance $d$ from $S$ along its
$x$-axis, with $d\gtrsim 1$ ly.
With these initial conditions, $S$ and $S'$ belong to the same plane of
simultaneity, and their $t$-coordinates are the same, $t_1=t'_1$.

Now, suppose that frame $S'$ starts to accelerate with constant acceleration
$\pm a$ along the $x$-axis of frame $S$ and maintains that acceleration until
time $t_2$, with final velocity $a(t_2-t_1)\ll c$. At that point, the system
$S'$ is a Lorentz system moving at a constant speed and, according to
Equation (\ref{eq0}), the instant $t'_2$ of $S'$ {\em simultaneous} with instant
$t_2$ of $S$ is now given by
\begin{equation}
t'_2=\frac{t_2-\frac{(\pm a)(t_2-t_1)d}{c^2}}{\sqrt{1-\frac{a^2(t_2-t_1)^2}{c^2}}}
\approx t_2-\frac{(\pm a)(t_2-t_1)d}{c^2}, \quad \textrm{for}\,\,\,
\frac{a(t_2-t_1)}{c}\ll 1.
\label{eq1}
\end{equation}

In Equation (\ref{eq1}), we neglect terms containing the 2nd power of
$v/c$. At time $t_2$, the position of $S'$ relative to frame $S$ is no more
$d$ but $d\pm\frac{1}{2}a(t_2-t_1)^2$, but we neglect that because
$\frac{1}{2}a(t_2-t_1)^2\ll d$.

Notice that, although the relative velocity $a(t_2-t_1)$ is much
less than $c$, the simultaneity term $\frac{a(t_2-t_1)d}{c^2}$ in
Equation (\ref{eq1}) is not negligible because of the large distance $d$.

Equation (\ref{eq1}) tells us that while for the inertial observer in $S$, an
interval of time $\Delta t =t_2-t_1$ has passed, the corresponding
interval of time $\Delta t'=t'_2-t'_1$ elapsed in reference frame $S'$ is
\begin{equation}
 \Delta t'=\Delta t\left[1-\frac{(\pm a)d}{c^2}\right].
\label{eq2}
\end{equation}

It is worth noticing that the application of Lorentz transformations to
accelerated frames is a straightforward practice {{\cite{des}}}.
For instance, it has been used to provide a solution to the Bell spaceship
paradox \cite{fra}.
Even Einstein used it in 1905 to derive time dilation for a clock moving
in a polygonal or continuously curved line \cite{e05}.
In fact, in Section \ref{se3}, we provide a derivation of our time
dilation/contraction formula that makes use of and generalizes Einstein's
derivation by extending it to systems subject to constant acceleration for a
short period of time. In the same section, a Minkowski diagram that visualizes
the origin of the effect is also given.

Now, suppose that during interval $\Delta t'$, the light source at rest in
$S'$ emits a beam of light of frequency $\nu'$. That means that $N$ wave
crests are emitted with $N=\nu'\Delta t'$. The same number of crests
must then be received by the observer in $S$ exactly after the traveling
time $d/c$, no matter how big $d/c$ is.
Moreover, the observer in $S$ will receive the $N$ wave crests within the
shorter interval of time $\Delta t$ because, for $S$, the whole emission
process in $S'$ has taken place within $\Delta t$ (the traveling time $d/c$
cannot affect that duration since $d/c$ is only a delay in receiving the
wave train).
That means that the observer in $S$ receives a beam of light of frequency $\nu$
such that $\nu\Delta t = N =\nu'\Delta t'$, and thus 
\begin{equation}
 \nu=\nu'\left[1-\frac{(\pm a)d}{c^2}\right].
\label{eq3}
\end{equation}

{{It is worth mentioning that Equations (\ref{eq2}) and (\ref{eq3})
correspond to those derived from general relativity in the approximation
of a weak and spatially uniform gravitational field $\pm a$. In fact,
within general relativity, they are obtained by using special relativity and the
principle of equivalence.}}

It is useful for the subsequent derivations to define the dimensionless
quantity $z$ as follows

\begin{equation}
 z\equiv \frac{\nu - \nu'}{\nu'}= -\frac{\pm ad}{c^2}.
\label{eq4}
\end{equation}

It is interesting to note that Equations (\ref{eq2}) and (\ref{eq3}) also hold if
the source $S'$ is inertial and stationary, and the observer $S$ accelerates
with acceleration $\pm a$.

As before, $S$ and $S'$ are initially inertial and relatively at rest, and
thus $t_1=t'_1$. Then, frame $S$ accelerates with acceleration $\pm a$ in the
$x$-direction until the instant $t_2$ and afterward moves at constant
velocity $\pm a(t_2-t_1)$. From this moment onward, it does not matter which
frame has accelerated and which is actually moving ($S$ with velocity
$\pm a(t_2-t_1)$
or $S'$ with velocity $\mp a(t_2-t_1)$). The Lorentz transformations are
`memory-less': in them, there is no mathematical dependence on the past
motion history of the reference frames. Then, the relations that give $t_2'$,
$\Delta t'$, and $\nu$ for $S$ are the same as Equations (\ref{eq1})--(\ref{eq3}).
We shall see later that this result has interesting philosophical
consequences.

In the present and the following section, we have also adopted the same
assumption made by Einstein in \cite{e07}, namely that acceleration $a$ has
negligible physical effects on the rate of clocks in the accelerated frame.
That is known as the `clock hypothesis' \cite{clo}.

\section{Detailed Derivation of the Time Dilation/Con\-trac\-tion Formula}
\label{se3}

Here, we take Einstein's derivation of the time dilation formula for a
clock moving in arbitrary motion (clock moving in a polygonal or continuously
curved line \cite{e05}) and apply it to the case of a system moving on a
straight line but subject to a uniform acceleration $a$ for a short period of
time. Hereafter, without loss of generality,
we assume that all the involved velocities are such that $v\ll c$.
We shall see that when acceleration $a$ goes to zero, one recovers the
well-known Einstein's time dilation formula. On the other hand, if the distance
between the inertial observer and the accelerating system is suitably large,
one recovers Equation (\ref{eq2}) of Section \ref{se2}.

Consistently with the previous section, primed quantities refer to the moving
system $S'$, while non-primed ones refer to the inertial (stationary) system
$S$. Moreover, $S'$ moves in the positive $x$-direction of $S$, and all
three coordinate axes are parallel. Suppose that $S'$ initially moves with
constant velocity $v_1$, and at time $t=t'=0$ the origins of $S$ and $S'$
overlap. Thus, the relation between the instants of time $t_1'$ of $S'$ and
$t_1$ of $S$ is given by Equation (\ref{eq0}) as follows

\begin{equation}
 t_1'=\frac{t_1 - \frac{v_1(v_1t_1)}{c^2}}{\sqrt{1-\frac{v_1^2}{c^2}}},
\label{eq5}
\end{equation}
since $x_1=v_1t_1$.

At instant $t_1$, the system $S'$ starts to accelerate in the positive or
negative $x$-direction with constant acceleration $a$, and at instant $t_2$
returns to uniform motion with the new constant velocity $v_2=v_1\pm
a(t_2-t_1)$.

Thus, the relation between the instants of time $t_2'$ of $S'$ and $t_2$ of
$S$ is now given by

\begin{equation}
 t_2'=\frac{t_2 - \frac{[v_1\pm a(t_2-t_1)][v_1t_1 + v_1(t_2-t_1) \pm
 \frac{1}{2}a(t_2-t_1)^2]}{c^2}}{\sqrt{1-\frac{[v_1\pm
 a(t_2-t_1)]^2}{c^2}}},
\label{eq6}
\end{equation}
where $x_2=v_1t_1 + v_1(t_2-t_1)\pm \frac{1}{2}a(t_2-t_1)^2$.

The interval of time $\Delta t'=t_2'-t_1'$ is thus equal to

\begin{equation}
 \Delta t'= \frac{t_2 - \frac{[v_1\pm a(t_2-t_1)][v_1t_1 + v_1(t_2-t_1)\pm
 \frac{1}{2}a(t_2-t_1)^2]}{c^2}}{\sqrt{1-\frac{[v_1\pm
 a(t_2-t_1)]^2}{c^2}}} - \frac{t_1 -
 \frac{v_1(v_1t_1)}{c^2}}{\sqrt{1-\frac{v_1^2}{c^2}}}.
\label{eq7}
\end{equation}

Now, it is not difficult to see that if we set $a=0$ in Equation (\ref{eq7})
and do not neglect terms containing the 2nd power of $v/c$, we recover
Einstein's time dilation formula

\begin{equation}
 \Delta t'= \Delta t \sqrt{1-\frac{v_1^2}{c^2}}.
 \label{eq8}
\end{equation}

On the other hand, if we set $v_1t_1=d$, with $d$ equal to an extremely large
astronomical distance, and if we consequently adopt the natural
approximations, $v_1(t_2-t_1)\pm \frac{1}{2}a(t_2-t_1)^2\ll v_1t_1$ and 
$\frac{[v_1\pm a(t_2-t_1)]^2}{c^2}\approx\frac{v_1^2}{c^2}\approx 0$ (we are
now neglecting again terms containing the 2nd power of $v/c$),
from Equation (\ref{eq7}) we arrive at the following relation

\begin{equation}
 \Delta t'= \Delta t \left[1-\frac{\pm ad}{c^2} \right],
\label{eq9}
\end{equation}
which is equal to Equation (\ref{eq2}).

In short, we have replicated Einstein's derivation of the time dilation for a
clock arbitrarily moving relative to a stationary clock \cite{e05}. Like
Einstein, we started from the Lorentz transformation of the time coordinate.
However, we have plugged in the equation an explicit and simpler type of motion
for the moving clock: namely, the moving clock moves away from the stationary
one on a straight line at constant velocity $v_1$ for a time $t_1$, and then,
for a time $(t_2-t_1)$, it accelerates with a low acceleration $a$. That is
simpler than Einstein's motion in a polygonal or continuously curved
line \cite{e05}. Therefore, if special relativity holds for non-uniform motion
in a ``continuously curved line'', it does also hold for a body slightly
accelerating in a straight line. By the way, what we have done so far is
equivalent to mapping the considered set-up onto a continuous sequence of
events that are analyzed relative to instantaneous co-moving inertial
frames.

In the remaining, we shall visualize the case with positive $a$ on a Minkowski
diagram. To make the graph easily readable, we shall assume, along with the
previous approximation $v/c\ll 1$, that the initial velocity $v_1$ of system
$S'$ is equal to zero, like in the case described in Section \ref{se2}. Thus,
until time $t_1$ (=$t_1')$, systems $S$ and $S'$ are
relatively at rest. Then, from $t_1$ to $t_2$, system $S'$ accelerates from
$0$ to final velocity $a(t_2-t_1)$, as seen from stationary system $S$.

\begin{figure}[h]
\hspace{-6pt}
\includegraphics[width=10cm]{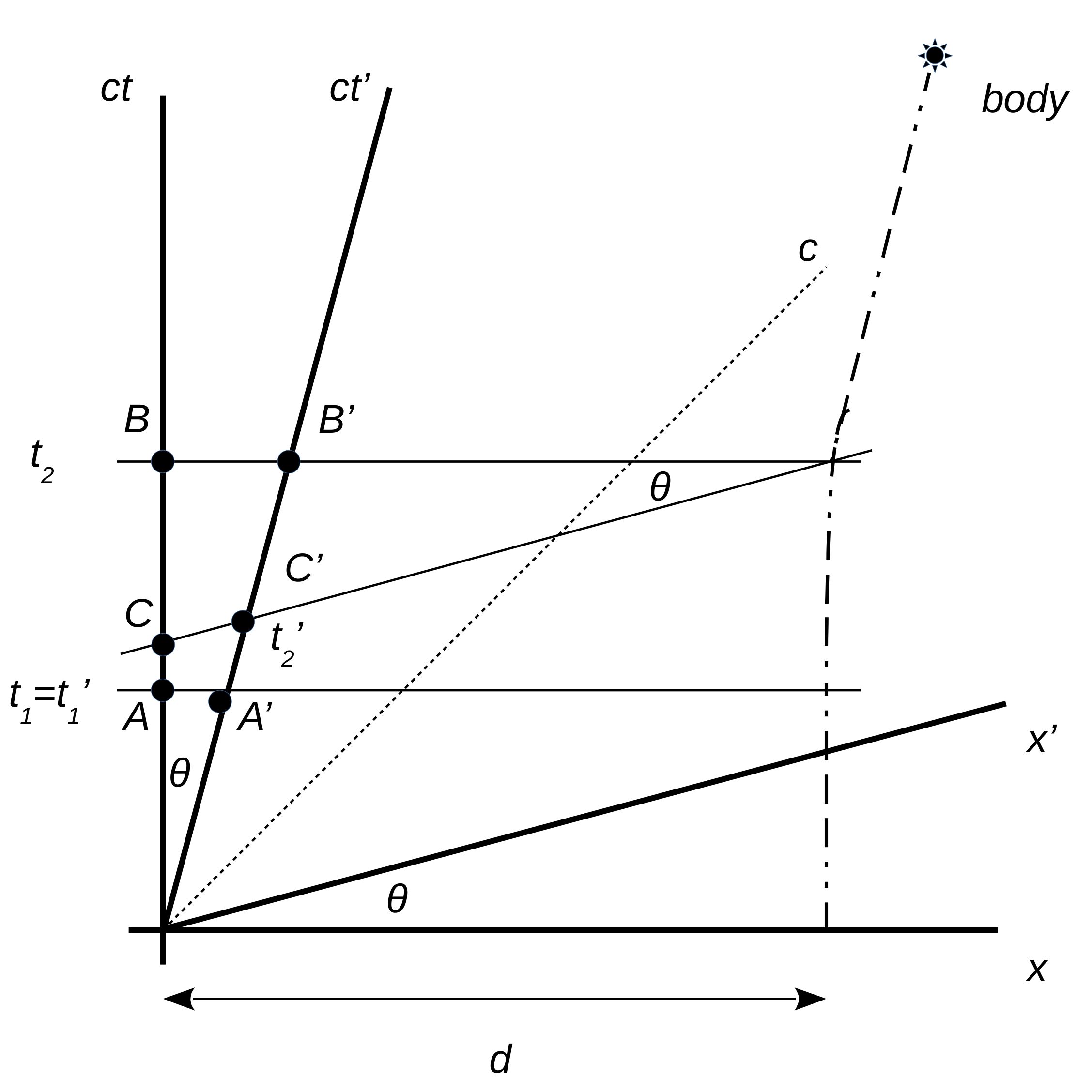}
\caption{Minkowski diagram of a body at astronomical distance $d$ from
 a stationary observer (system $x$-$ct$). The body is at rest until time
 $t_1$ $(=t_1')$ and then accelerates to the final constant speed $a(t_2-t_1)$
 in $\Delta t =t_2-t_1$. The dot-dashed line is the world line of the body.
 Quantities are not to scale. The magnitude of angle $\theta$ has been
 exaggerated for representation purposes.}
\label{fig1}
\end{figure}

As shown in Figure \ref{fig1}, the dot-dashed line represents the world line
of a body placed at an astronomical distance $d$ in the reference frame $S'$.
Its world line is initially vertical because, until time $t_1$ $(=t_1')$, both
systems are relatively at rest. Then, between time $t_1$ and $t_2$, system $S'$
and the body in it accelerate from $0$ to final velocity $a(t_2-t_1)$.

Within $\Delta t=t_2-t_1$, the body's world line gets bent to an angle $\theta$
relative to the $ct$ axis such that
$\tan\theta=\frac{v}{c}=\frac{a(t_2-t_1)}{c}$,
and then it stays parallel to the $ct'$ axis of $S'$.
Since $\frac{v}{c}=\frac{a(t_2-t_1)}{c}\ll 1$, neglecting 2nd order terms in
$v/c$, the unit lengths $U$ of space-time axes of $S$ and $S'$ can be taken
as identical,
\begin{equation}
 U'= U\sqrt{\frac{1+\frac{v^2}{c^2}}{1-\frac{v^2}{c^2}}}\approx U.
 \label{eq10}
\end{equation}

With reference to Figure \ref{fig1}, $\overline{BA}=ct_2 -ct_1=c\Delta t$ and
$\overline{C'A'}=ct_2' -ct_1'=c\Delta t'$.

Since $\frac{v}{c}\ll 1$ and
$\theta\approx 0$, we have that $\overline{C'A'}\approx \overline{CA}$ and
$\overline{C'B'}\approx \overline{CB}$. Then,
$c\Delta t -c\Delta t' = \overline{BA}-\overline{C'A'}\approx\overline{BA}-
\overline{CA}=\overline{CB}=d\tan\theta =
\frac{vd}{c}=\frac{a(t_2-t_1)d}{c}=\frac{a\Delta td}{c}$.
Thus,
\begin{equation}
 c\Delta t -c\Delta t' =\frac{a\Delta td}{c}\quad\to\quad \Delta t'=\Delta t
 \left[1-\frac{ad}{c^2}\right].
 \label{eq11}
\end{equation}

\vspace{-6pt}

\section{Proxima Centauri}
\label{se4}

We now apply Equation (\ref{eq4}) to the light emitted by Proxima Centauri at
distance $d=4.25\,$ly ($d\simeq 4.02\times 10^{16}\,$m) to estimate the expected
$z$ and compare it with the standard Doppler shift ascribable to the radial
velocity of the star imparted by orbiting planet Proxima Centauri b.

In the scheme of Section \ref{se2}, we assume Proxima Centauri to correspond
to frame $S'$, while we, the observers, are stationary in inertial frame $S$.
Moreover, for the sake of derivation, we assume that we are on the orbital
plane of Proxima b, and the acceleration of the star is along the line of
sight. 

Within that approximation and assuming a circular orbit, the projected
position $r(t)$ of Proxima b relative to Proxima Centauri can be written as
$r(t)=r_0\cos (\omega t)$, where $r_0=7.3\times 10^9\,$m is the planet
semi-major axis and $\omega=\frac{2\pi}{T}$ with orbital period
$T=11.18\,\textrm{d}=9.7\times 10^5\,$s. The mass of Proxima b is estimated to
be $M_{Pr\,b}=9.6\times 10^{24}\,$kg, and the mass of Proxima Centauri is
estimated to be $M_{Pr\,C}=2.4\times 10^{29}\,$kg. All the astronomical data are
from \cite{pro}.

The maximum value of Proxima b acceleration is given by
$a_{Pr\,b}=|\ddot r(t)|_{max}$ $=r_0\omega^2\simeq 0.306\,$m/s$^2$. The maximum
value of Proxima Centauri acceleration is then obtained by applying the third
law of dynamics, $M_{Pr\, C}\cdot a_{Pr\, C} = m_{Pr\, b}\cdot a_{Pr\,b}$, giving
$ a_{Pr\, C} =\frac{ m_{Pr\, b}}{M_{Pr\, C}} a_{Pr\,b}\simeq
1.21\times 10^{-5}\,$m/s$^2$.

Thus, the maximum absolute value of $z$ is

\begin{equation}
 |z|_{max}=\left|\frac{\nu - \nu'}{\nu'}\right|=
 \frac{a_{Pr\, C}\cdot d}{c^2}\simeq
 5.4\times 10^{-6}.
\label{eq12}
\end{equation}

Let us compare this value with the maximum radial
Doppler shift within the same approximation. The maximum value of Proxima b
radial velocity is given by $v_{Pr\,b}=|\dot r(t)|_{max}$ $=r_0\omega\simeq
4.7\times 10^4\,$m/s. The maximum value of Proxima Centauri radial velocity
is then obtained by applying the conservation of linear momentum,
$M_{Pr\, C}\cdot v_{Pr\, C} = m_{Pr\, b}\cdot v_{Pr\,b}$, giving
$ v_{Pr\, C} =\frac{ m_{Pr\, b}}{M_{Pr\, C}} a_{Pr\,b}\simeq 1.9\,$m/s.

Thus, the maximum absolute value of $z_{Doppler}=\pm\frac{v}{c}$ (for $v\ll c$)
due to the standard Doppler shift is

\begin{equation}
 |z|_{max\, Doppler}=\left|\frac{\nu - \nu'}{\nu'}\right|=
 \frac{v_{Pr\, C}}{c}\simeq 7.7\times 10^{-9}.
\label{eq13}
\end{equation}

\section{Discussion}
\label{se5}

In Sections \ref{se2} and \ref{se3}, we have shown that the derivation of
Equations (\ref{eq2}) and (\ref{eq3}) is straightforward and sound.
However, as anticipated by the calculations in the preceding section, these
new relations bring several issues with themselves that we shall discuss.

First and foremost, with the relatively close Proxima Centauri, the
maximum frequency shift due to Equation (\ref{eq3}) is expected to be three orders
of magnitude larger than the maximum Doppler shift due to its radial velocity
(Equations (\ref{eq12}) and (\ref{eq13})). Unfortunately, no such phenomenon has
been found in any observational data so far.

Moreover, as the distance $d$ between the emitting source and the Earth
increases, the frequency shift should become more and more dramatic, let alone
the fact that for suitably large $d$, we could have {\em negative
}
frequency $\nu$ and $\Delta t'$. As far as this author knows,
that has no immediate physical meaning.

A further problem comes from the case in which the light source
always remains stationary in an inertial frame (frame $S'$ inertial) while the
observer accelerates (frame $S$ accelerating).
As we have already shown at the end of Section \ref{se2},
Equations (\ref{eq2})--(\ref{eq4}) are still applicable to this case.
Thus, suppose we are in that situation and now receive a light signal
emitted with frequency $\nu'$ by a source stationary in an inertial
frame distant $d=$ 6 billion ly from us. Therefore, the signal was emitted 6
billion years ago. The point is: what would
be the frequency $\nu$ of the light signal that we detect now? According
to Equation (\ref{eq3}), the frequency $\nu$ also depends upon our acceleration $a$
relative to the emitter at the epoch of the emission, then the
actual value of $\nu$ is doomed to remain indeterminate. Six billion years
ago, we did not exist as observers, not to mention the state of motion of
our reference frame relative to the source back then. We have no doubt, though,
that we do receive a definite frequency.

How can all this be settled? This last conundrum suggests that the problem may
reside in the simultaneity term $-\frac{vx}{c^2}$ of the time coordinate
transformation (\ref{eq0}), particularly its dependence upon distance $x$.

In the end, our findings appear to be yet another relativity paradox.
As usually happens with special relativity, every new paradoxical result,
provided that it is formally correct, is always considered physically
real. It is considered an inescapable consequence of special relativity, however
counter-intuitive and lacking experimental confirmation may be.

With the present case, though, we believe there is something different going on.
Here, we have macroscopic proofs (on the astronomical scale) that something is
not working as expected in the machinery of special relativity, not from
a mathematical but a physical point of view.
We have no simple solution to this paradox. However, we believe that the
problem is in itself real and worth to be described and discussing.

\section*{Funding}
This research received no external funding.

\section*{Institutional Review Board Statement}
Not applicable.

\section*{Informed Consent Statement}
Not applicable.

\section*{Data Availability Statement}
Not applicable.

\section*{Acknowledgments}
The author is indebted to Assunta Tataranni and Gianpietro Summa for key
improvements to the manuscript. We thank two anonymous reviewers whose
comments/suggestions helped improve and clarify this manuscript.

\section*{Conflict of Interest}
The author declares no conflict of interest.

\end{document}